\documentclass[aps,prd,onecolumn,groupedaddress,showpacs,showkeys,nofootinbib]{revtex4}
\usepackage{epsfig} \usepackage{amsmath} \usepackage{amsfonts}
\usepackage{amssymb} \usepackage{graphicx} \usepackage{colordvi}
\begin{document} \title{The post-Minkowskian limit of
$f(R)$-gravity}

\author{Salvatore Capozziello$^{\diamond}$,\,Arturo Stabile$^{\natural}$, \,Antonio
Troisi$^{\ddag}$}

\address{$^{\diamond}$ Dipartimento di Scienze Fisiche, Universit\`a di Napoli "Federico II" and INFN sez. di Napoli
Compl. Univ. di Monte S. Angelo, Edificio
G, Via Cinthia, I-80126 - Napoli, Italy}

\address{$^{\natural}$ Dipartimento di Ingegneria, Universita' del Sannio and INFN sez. di Napoli, gruppo collegato di Salerno,
C-so Garibaldi, I - 80125 Benevento, Italy}

\address{$^{\ddag}$ Dipartimento di Ingegneria Meccanica, Universita'  di Salerno,
Via Ponte Don Melillo, I - 84084 Fisciano (SA), Italy}

 \date{\today}
%----------------------------------------------------------------
\begin{abstract}
We formally discuss  the post-Minkowskian limit of
 $f(R)$-gravity without adopting conformal transformations but developing all the calculations in the original Jordan frame. It is shown that such an approach
 gives rise, in general,  together with  the standard massless graviton, to  massive scalar modes whose masses are directly related to the analytic parameters of the theory.  In this sense, the presence of massless gravitons only is  a peculiar feature of General Relativity. This fact is never stressed enough and could have dramatic consequences in detection of gravitational waves. Finally the role of curvature stress-energy tensor of $f(R)$-gravity is discussed showing that it generalizes the so called Landau-Lifshitz tensor of General Relativity. The further degrees of freedom, giving rise to the massive modes, are directly related to the structure of such a tensor.

\end{abstract}


\maketitle
%----------------------------------------------------------------
%\renewcommand{\theequation}{\thesection.\arabic{equation}}
%----------------------------------------------------------------
\section{Introduction}
Astrophysical   observations of the last
decade suggests  the introduction
of new ingredients  in order to achieve a self-consistent picture of cosmos.   In particular, the observation
that Hubble flow is currently experiencing a  speeding up
has completely changed the approach to standard cosmology inducing to take
into account theoretical approaches more general than the standard lore of General Relativity (GR).

The simplest explanation of  such a cosmic acceleration requires to include the cosmological constant  in the
Friedmann-Robertson-Walker cosmology  (Concordance Model).  This ingredient  gives rise to a
negative pressure contribution needed to balance the standard matter
attractive interaction. Although the Concordance Model represents
the best fit model with respect  to all samples of data coming from supernovae, large-scale structure and cosmic microwave radiation
\cite{LCDM}, several  conceptual problems come out to  theoretically define and  to give an explanation to  the observed value of the cosmological constant . 
Furthermore, assuming the existence of both dark energy and dark matter, we should find out new fundamental ingredients capable of giving account to almost 
95$\%$ of the total amount of cosmic matter-energy. 

Due to these difficulties, 
people have considered alternative approaches to GR that could be able to
 frame the observed late time acceleration  and missing matter without introducing new ingredients. 
In this sense, higher order gravity \cite{HOG} and, in
particular, fourth order gravity  represent an
interesting scheme which could, potentially, address the problems. 

Up to now, these theories have been
investigated both at  cosmological scale  and in the weak field limit
with significant results \cite{f(R)-noi,f(R)-cosmo,zerbini,weakf(R)}. It
has been shown that an accelerating late time behaviour can be
easily recovered  \cite{saffari} and, in addition, it can be
coherently related to an early time inflationary expansion
\cite{odintsov07}. Furthermore, such an approach seems to deserve
attention even at smaller scales. In fact, modifying the gravity
action in favor of a non-linear Lagrangian in the Ricci scalar
implies, in the Newtonian limit, corrections to the gravitational
potential which can induce an astrophysical phenomenology
interesting at galactic scales.  In particular one can fit the
rotation curves of spiral galaxies and the haloes of galactic clusters without  the
introduction of dark matter \cite{noi-mnras}. 
Besides, several of these extended models evade Solar System tests so they are not in conflict with positive experimental results of GR  \cite{ pogosian,tsujikawa}.

A relevant aspect of higher order gravity theories is that,
in the post-Minkowskian limit (i.e. small fields and no
prescriptions on the propagation velocity), the propagation of the
gravitational fields turns out to be characterized by  waves
 with both  tensorial and scalar modes \cite{stelle78,greci}.
This issue represents a  striking difference between  GR
 and extended gravity  since, in the standard Einstein
scheme, only tensorial degrees of freedom are allowed. As matter of
facts, the gravitational waves can represent a fundamental tool to
 discriminate between GR and
alternative gravities \cite{stochastic, will}.  

In this paper, we want to develop, formally,  the post-Minkowkian limit of analytic  $f(R)$-gravity models which, in our opinion, has never been pursued with accuracy stressing enough some peculiar points.  As shown  by the same authors for the Newtonian limit, we will show that it is different from the same limit of GR since massive modes naturally come out in the gravitational radiation. This occurrence has a deep meaning since points out that the presence of massless modes only  is nothing else but the  particular case of GR while massive and ghost modes are present in general \cite{greci}.

The layout of the paper is the following.  In Sec. \ref{2}, we discuss the post-Minkowskian limit of $f(R)$-gravity. Considerations on gravitational wave massive modes are developed in Sec.\ref{3}. Sec.\ref{4} is devoted to the discussion of the role of the stress-energy tensor in $f(R)$-gravity. Concluding remarks are drawn in Sec. \ref{5}.

\section{The post-Minkowskian limit of $f(R)$ - gravity}
\label{2}
Any theory of gravity has to be discussed in the weak field limit approximation.  This  "prescription" is needed to test if the  given theory is consistent with the well-established Newtonian Theory and with the Special Relativity as soon as the the gravitational field is weak or is almost null. Both requirements are fulfilled by GR and then they can be considered  two possible paradigms to confront a given theory, at least in the weak field limit, with GR itself.  
In \cite{noi-newt,dirk},  the Newtonian limit of $f(R)$-gravity is investigated always remaining in the Jordan frame \cite{faragunz}. From our point of view, this is important since, by perturbatively approximating a field, some conformal features could be lost. Here we want to derive, formally, the post-Minkowskian limit of $f(R)$-gravity.
 
 The post-Minkowskian limit of any theory of  gravity arises when the
regime of small field is considered without any prescription on
 the propagation  of the field. This case has to be
clearly distinguished with respect to the Newtonian limit which,
differently, requires both the small velocity and the weak field
approximations. Often, in literature, such a distinction is not
clearly remarked and several cases of pathological analysis can be
accounted. The post-Minkowskian limit of GR gives rise to
massless  gravitational waves. An analogous study can be pursued considering, instead of 
 the Hilbert-Einstein Lagrangian linear in the Ricci scalar $R$,  a general function $f(R)$. The only assumption that we are going to do is that $f(R)$ is an analytic function. The gravitational action is then
\begin{equation}
{\cal A}=\int
d^4x\sqrt{-g}\biggl[f(R)+\mathcal{X}\mathcal{L}_m\biggr]\,,
\end{equation}
where ${\displaystyle \mathcal{X}=\frac{16\pi G}{c^4}}$ is the coupling,   $\mathcal{L}_m$ is the
standard matter Lagrangian and $g$ is the determinant of the metric.
The field equations, in metric formalism, read\footnote{All considerations are developed here in metric formalism.}

\begin{equation}\label{fe1}
f'R_{\mu\nu}-\frac{1}{2}fg_{\mu\nu}-f'_{;\mu\nu}+g_{\mu\nu}\Box_g
f'=\frac{\mathcal{X}}{2}T_{\mu\nu}
\end{equation}
\begin{equation}\label{TrHOEQ}
3\Box f'+f'R-2f\,=\,\frac{\mathcal{X}}{2}T\,,
\end{equation}
with
${\displaystyle T_{\mu\nu}=\frac{-2}{\sqrt{-g}}\frac{\delta(\sqrt{-g}\mathcal{L}_m)}{\delta
g^{\mu\nu}}}$ the energy momentum tensor of matter ($T$ is the
trace), ${\displaystyle f'=\frac{df(R)}{dR}}$ and
$\Box_g={{}_{;\sigma}}^{;\sigma}$. We adopt a $(+,-,-,-)$
signature, while the conventions for Ricci's tensor is
$R_{\mu\nu}={R^\sigma}_{\mu\sigma\nu}$ and 
${R^\alpha}_{\beta\mu\nu}=\Gamma^\alpha_{\beta\nu,\mu}+...$ for the Riemann tensor, where

\begin{equation}\label{chri}
\Gamma^\mu_{\alpha\beta}=\frac{1}{2}g^{\mu\sigma}(g_{\alpha\sigma,\beta}+g_{\beta\sigma,\alpha}-g_{\alpha\beta,\sigma})
\end{equation}
are the Christoffel symbols  of the $g_{\mu\nu}$
metric. Actually, in order to perform a post-Minkowskian
limit of field equations, one has to perturb  Eqs.
(\ref{fe1})  on the Minkowski background
$\eta_{\mu\nu}$. In such a case  the invariant metric element  becomes

\begin{equation}\label{me} ds^2=g_{\sigma\tau}dx^\sigma
dx^\tau=(\eta_{\sigma\tau}+h_{\sigma\tau})dx^\sigma dx^\tau
\end{equation}
with $h_{\mu\nu}$ small ($O(h)^2\ll 1$). 
We assume that the $f(R)$-Lagrangian is  analytic
 (i.e. Taylor expandable) in term of the Ricci scalar, which means that 
 
\begin{eqnarray}\label{sertay}
f(R)=\sum_{n}\frac{f^n(R_0)}{n!}(R-R_0)^n\simeq
f_0+f'_0R+\frac{1}{2}f''_0R^2+...\,.
\end{eqnarray}
The flat-Minkowski background is recovered for   $R=R_0\simeq 0$.

Field equations (\ref{fe1}), at the first order of
approximation in term of the perturbation \cite{spher},
become\,:

\begin{equation}\label{fe2}
f_0'\biggl[R^{(1)}_{\mu\nu}-\frac{R^{(1)}}{2}\eta_{\mu\nu}\biggr]-f''_{0}\biggl[R^{(1)}_{,\mu\nu}-\eta_{\mu\nu}\Box
R^{(1)}\biggr]=\frac{\mathcal{X}}{2}T^{(0)}_{\mu\nu}
\end{equation}
where ${\displaystyle f'_0=\frac{df}{dR}\Bigl|_{R=0}}$,
${\displaystyle f''_0=\frac{d^2f}{dR^2}\Bigl|_{R=0}}$ and
$\Box={{}_{,\sigma}}^{,\sigma}$ that is now the standard d'Alembert operator of flat space-time. 
From the zero-order of Eqs.(\ref{fe1}),
one gets $f(0)=0$, while $T_{\mu\nu}$ is fixed at zero-order in
Eq.(\ref{fe2}) since, in this perturbation scheme, the first order on
Minkowski space has to be connected with the zero order of the
standard matter energy momentum tensor\footnote{This formalism
descends from the theoretical setting of Newtonian mechanics which
requires the appropriate scheme of approximation when obtained
from a more general relativistic theory. This scheme coincides
with a gravity theory analyzed at the first order of perturbation
in the curved spacetime metric.}. The explicit expressions of
the Ricci tensor and scalar, at the first order in the metric
perturbation, read

\begin{equation}\label{approx1}
\left\{\begin{array}{ll}R^{(1)}_{\mu\nu}=h^\sigma_{(\mu,\nu)\sigma}-\frac{1}{2}\Box
h_{\mu\nu}-\frac{1}{2}h_{,\mu\nu}\\\\
R^{(1)}={h_{\sigma\tau}}^{,\sigma\tau}-\Box h \end{array}\right.
\end{equation}
with $h={h^\sigma}_\sigma$. Eqs. (\ref{fe2}) can be written in a
more suitable form by introducing the constant
${\displaystyle \xi=-\frac{f''_0}{f'_0}}$, that is

\begin{equation}\label{fe3}
h^\sigma_{(\mu,\nu)\sigma}-\frac{1}{2}\Box
h_{\mu\nu}-\frac{1}{2}h_{,\mu\nu}-\frac{1}{2}({h_{\sigma\tau}}^{,\sigma\tau}-\Box
h)\eta_{\mu\nu}+\xi(\partial^2_{\mu\nu}-\eta_{\mu\nu}\Box)({h_{\sigma\tau}}^{,\sigma\tau}-\Box
h)=\frac{\mathcal{X}}{2f'_0}T^{(0)}_{\mu\nu}\,.
\end{equation}
By choosing the transformation
$\tilde{h}_{\mu\nu}=h_{\mu\nu}-\frac{h}{2}\eta_{\mu\nu}$ and the
gauge condition $\tilde{h}^{\mu\nu}_{\,\,\,\,\,\,\,,\mu}=0$, one
obtains that field equations  and the trace equation, respectively, read
\footnote{The gauge transformation is $h'_{\mu\nu}=h_{\mu\nu}-\zeta_{\mu,\nu}-\zeta_{\nu,\mu}$ when we
perform a coordinate transformation as $x'^\mu=x^\mu+\zeta^\mu$
with O($\zeta^2$)$\ll 1$. To obtain the gauge and the validity of
the field equations for both perturbation $h_{\mu\nu}$ and
$\tilde{h}_{\mu\nu}$, the $\zeta_\mu$ have to satisfy the harmonic
condition $\Box\zeta^\mu=0$.}

\begin{equation}\label{fe4}
\left\{\begin{array}{ll}\Box\tilde{h}_{\mu\nu}+\xi(\eta_{\mu\nu}\Box-\partial^2_{\mu\nu})\Box\tilde{h}=-\frac{\mathcal{X}}{f'_0}T^{(0)}_{\mu\nu}
\\\\
\Box\tilde{h}+3\xi\Box^2\tilde{h}=-\frac{\mathcal{X}}{f'_0}T^{(0)}\end{array}
\right.\,.
\end{equation}
In order to derive the analytic solutions of Eqs. (\ref{fe4}), we can
adopt a momentum- description.  This approach
 simplifies the equations  and allows to
fix the physical properties of the problem.
In such a scheme, we have\,:
\begin{equation}\label{femomentum}
\left\{\begin{array}{ll}k^2\tilde{h}_{\mu\nu}(k)+\xi(k_\mu
k_\nu-k^2\eta_{\mu\nu})k^2\tilde{h}(k)=\frac{\mathcal{X}}{f'_0}T^{(0)}_{\mu\nu}(k)
\\\\ k^2\tilde{h}(k)(1-3\xi
k^2)=\frac{\mathcal{X}}{f'_0}T^{(0)}(k)\end{array} \right.
\end{equation}
where

\begin{equation}\label{solgen}
\left\{\begin{array}{ll}\tilde{h}_{\mu\nu}(k)=\int\frac{d^4x}{(2\pi)^2}\tilde{h}_{\mu\nu}(x)\,\,e^{-ikx}
\\\\
T^{(0)}_{\mu\nu}(k)=\int\frac{d^4x}{(2\pi)^2}T^{(0)}_{\mu\nu}(x)\,\,e^{-ikx}\end{array}
\right. \end{equation} are the Fourier transforms of the
perturbation $\tilde{h}_{\mu\nu}(x)$ and of the matter tensor
$T^{(0)}_{\mu\nu}$. We have defined, as usual, $k\,x=\omega
t-\mathbf{k}\cdot\mathbf{x}$ and $k^2=\omega^2-\mathbf{k}^2$; 
 $\tilde{h}(k)$ and $T^{(0)}(k)$ are the traces of
$\tilde{h}_{\mu\nu}(k)$ and $T^{(0)}_{\mu\nu}(k)$. In the momentum
space, one can easily recognize the solutions of
Eqs.(\ref{femomentum});  $\tilde{h}_{\mu\nu}(k)$
turns out to be

\begin{eqnarray}\label{x}\tilde{h}_{\mu\nu}(k)=\frac{\mathcal{X}}{f'_0}\frac{T^{(0)}_{\mu\nu}(k)}{k^2}+\frac{\mathcal{\xi
X}}{f'_0}\frac{k^2\eta_{\mu\nu}-k_\mu k_\nu}{k^2(1-3\xi
k^2)}T^{(0)}(k),
\end{eqnarray}
which fulfils the condition
$\tilde{h}^{\mu\nu}_{\,\,\,\,\,\,\,,\mu}=0$
(that is $\tilde{h}^{\mu\nu}(k)\,\,k_\mu=0$). The  perturbation
variable $h_{\mu\nu}(k)$ can be obtained by inverting the relation
with the tilded variables.  In particular, by inserting the new stress-energy tensor
$S^{(0)}_{\mu\nu}(k)=T^{(0)}_{\mu\nu}(k)-\frac{1}{2}\eta_{\mu\nu}T^{(0)}(k)$
with the trace $S^{(0)}(k)=\eta^{\mu\nu}S^{(0)}_{\mu\nu}(k)$, one obtains:

\begin{eqnarray}\label{y}h_{\mu\nu}(k)=\frac{\mathcal{X}}{f'_0}\frac{S^{(0)}_{\mu\nu}(k)}{k^2}+\frac{\mathcal{\xi
X}}{2f'_0}\frac{k^2\eta_{\mu\nu}+2k_\mu k_\nu}{k^2(1-3\xi
k^2)}S^{(0)}(k),
\end{eqnarray}
which represents a wave-like solution, in the momentum space, with
a massless and a massive contributions. The massive term is due to the pole in the
denominator of the second term: the {\it mass} is directly related
with the physical properties of the pole itself  and, thanks to the parameter $\xi$,  depends on the analytic form of the model (i.e. $f'_0$ and $f''_0$). The 
wavelike solution in  the configuration
space is obtained by  the inverse Fourier transform of $h_{\mu\nu}(k)$.

\section{Massive modes in gravitational waves}
\label{3}
The presence of the massive term is a feature emerging from the intrinsic non-linearity of $f(R)$-gravity. Specifically, it is related to the fact that  $f''_0\neq 0$, which is zero in GR where $f(R)=R$. This means that massless states are nothing else but a particular case among the gravitational theories. A similar situation emerges also in the Newtonian limit:  the Newton potential is recovered only as the weak field limit of GR. In general, Yukawa-like corrections, and then characteristic interaction lengths, are present \cite{noi-newt}

Some considerations are in order at this point. 
It is worth noticing that field Eqs. (\ref{fe1}) can be written putting in evidence the Einstein
tensor  in the l.h.s. \cite{f(R)-noi}. In such a case, higher than second order
differential contributions can be
considered  as a sources  in the r.h.s.  as well as the energy-momentum tensor of standard matter:

\begin{equation} \label{eq: field}
G_{\mu \nu} = R_{\mu \nu} - \frac{1}{2} R g_{\mu \nu} =
T^{(curv)}_{\mu \nu} + T^{(m)}_{\mu \nu}\,,
\end{equation}
where

\begin{equation}\label{eq: curvstress}\left\{\begin{array}{ll}T^{(m)}_{\mu\nu} =\frac{T_{\mu\nu}}{f'(R)} \\\\ T^{(curv)}_
{\mu\nu}
=\frac{1}{2}g_{\mu\nu}\frac{f(R)-f'(R)R}{f'(R)}+\frac{f'(R)_{;\mu\nu}-g_{\mu\nu}\Box_g
f'(R)}{f'(R)}\,.\end{array} \right.
\end{equation}
Actually, if we consider the perturbed metric (\ref{me}) and
develop the Einstein tensor up to the first order in perturbations, 
we have

\begin{equation}G_{\mu\nu}\sim G^{(1)}_{\mu\nu}=h^\sigma_{(\mu,\nu)\sigma}-\frac{1}{2}\Box
h_{\mu\nu}-\frac{1}{2}h_{,\mu\nu}-\frac{1}{2}({h_{\sigma\tau}}^{,\sigma\tau}-\Box
h)\eta_{\mu\nu}
\end{equation}
while the curvature stress-energy tensor  gives the  contributions

\begin{equation}T^{(curv)}_{\mu\nu}\sim\xi(\partial^2_{\mu\nu}-\eta_{\mu\nu}\Box)({h_{\sigma\tau}}^{,\sigma\tau}-\Box
h)\,.
\end{equation}
This expression allows  to recognize that, in the  space
of momenta, such a quantity will be responsible of the pole-like
term which implies the introduction of a massive degree of freedom
into the particle spectrum of gravity. In fact, inserting these
two expressions into the the field Eqs. (\ref{eq: field}) and
considering Eqs.(\ref{approx1}), we obtain the solution\,:

\begin{eqnarray}\Box
h_{\mu\nu}(x)=-\frac{\mathcal{X}}{f'_0}\biggl[S^{(0)}_{\mu\nu}(x)+\Sigma^\xi_{\mu\nu}(x)\biggr]
\end{eqnarray}
where $\Sigma^\xi_{\mu\nu}(x)$ is related to the curvature stress-energy
tensor and is defined as

\begin{eqnarray}\Sigma^\xi_{\mu\nu}(x)=\frac{\mathcal{\xi}}{2}\int\frac{d^4k}{(2\pi)^2}\frac{k^2\eta_{\mu\nu}+2k_\mu
k_\nu}{1-3\xi k^2}S^{(0)}(k)\,\,e^{ikx}\,.
\end{eqnarray}
The general solution for the metric perturbation $h_{\mu\nu}(x)$,
when the field equations are  (\ref{eq: field}), can be rewritten
as

\begin{eqnarray}\label{gensolFT}
h_{\mu\nu}(x)=\frac{\mathcal{X}}{f'_0}\int\frac{d^4k}{(2\pi)^2}\frac{S^{(0)}_{\mu\nu}(k)}{k^2}\,\,e^{ikx}+\frac{\mathcal{\xi
X}}{2f'_0}\int\frac{d^4k}{ (2\pi)^2}\frac{k^2\eta_{\mu\nu}+2k_\mu
k_\nu}{k^2(1-3\xi k^2)}S^{(0)}(k)\,\,e^{ikx}\,,
\end{eqnarray}
where   the second  pole-like term is present.  In vacuum (i.e.
$T^{(m)}_{\mu\nu}=0$), Eqs. (\ref{fe4}) become

\begin{equation}\label{fe5}
\left\{\begin{array}{ll}k^2[\tilde{h}_{\mu\nu}(k)+\xi(k_\mu
k_\nu-k^2\eta_{\mu\nu})\tilde{h}(k)]=0 \\\\ k^2\tilde{h}(k)(1-3\xi
k^2)=0\end{array} \right.
\end{equation}
showing that allowed solutions are of two types, i.e.\,:
\begin{equation}\label{sol1}
\left\{\begin{array}{ll} \omega=\pm |\mathbf{k}| \\\\
h_{\mu\nu}(x)=\int\frac{d^4k}{(2\pi)^2}h_{\mu\nu}(k)\,\,e^{ikx}&\,\,\,\,\text{with}
\,\,\,\,\,\,h(k)=0\end{array} \,,\right.
\end{equation}
and

\begin{equation}\label{sol2}
\left\{\begin{array}{ll} \omega=\pm
\sqrt{\mathbf{k}^2+\frac{1}{3\xi}} \\\\
h_{\mu\nu}(x)=-\int\frac{d^4k}{(2\pi)^2}\biggl[\frac{\eta_{\mu\nu}+6\xi
k_\mu k_\nu}{6}\biggr]\,h(k)\,\,e^{ikx}&\,\,\,\,\text{with}
\,\,\,\,\,\,h(k)\neq 0\end{array} \,.\right.
\end{equation}
It is evident, that the first solution represents a massless
graviton according to the standard prescriptions of GR while the
second one gives a massive degree of freedom with $m^2=
(3\xi)^{-1}=-\frac{f'_0}{3f''_0}$. Thanks to this propertiy, we
can  rewrite Eqs.(\ref{fe4}) introducing a scalar field
$\phi=\Box\tilde{h}$ so that the general system can be rearranged
in the following way

\begin{equation}
\left\{\begin{array}{ll}\Box\tilde{h}_{\mu\nu}=-\frac{\mathcal{X}}{f'_0}T^{(0)}_{\mu\nu}+\biggl[\frac{\partial^2_{\mu\nu}-\eta_{\mu\nu}\Box}{3m^2}\biggr]\phi
\\\\
(\Box+m^2)\phi=-\frac{\mathcal{X}}{f'_0}m^2T^{(0)}\end{array}
\right.
\end{equation}
which suggests that the higher order contributions act, in the
post-Minkowskian limit, as a massive scalar field whose mass
depends on the derivatives  $f'(R)$ and $ f''(R)$, calculated
on the unperturbed background  metric.

The massive mode is directly related to the coefficients of the 
Taylor expansion and it is interesting to note that they dermine also the value of the Yukawa correction in the Newtonian approximation \cite{noi-newt,dirk}. 
On the other hand,  it is straightforward to see that massive modes are directly related to the non-trivial structure of the trace equation as it is easy to see 
from   Eq.(\ref{TrHOEQ}). In GR,  the Ricci scalar is univocally fixed being $R=0$ in vacuum and $R\propto \rho$ in presence of matter, where $\rho$ is the matter-energy density.

\section{The stress-energy tensor in  $f(R)$-gravity and the gravitational radiation}
\label{4}
As we have seen, higher order theories of gravity introduce further degrees of fredom  which can be taken into account by defining an additional "curvature source term" in the r.h.s. of field equations.
This quantity behaves as an effective stress-energy tensor that
can  characterize the energy loss due to the gravitational
radiation. Although the procedure to calculate the
stress-energy tensor of the gravitational field in GR is often
debated, one can extend the formalism to more general
theories and obtain this quantity by varying  the
gravitational Lagrangian. In GR, this quantity is a pseudo-tensor  and  is tipically
referred to  as the Landau-Lifshitz energy-momentum tensor \cite{landau}.

The
calculations of GR need to be extended when dealing with  higher order
gravity.
In the case of $f(R)$-gravity, we have

\begin{eqnarray}
\delta\int d^4x\sqrt{-g}f(R)=\delta\int
d^4x\mathcal{L}(g_{\mu\nu},g_{\mu\nu,\rho},g_{\mu\nu,\rho\sigma})\approx\int
d^4x\biggl(\frac{
\partial\mathcal{L}}{\partial g_{\rho\sigma}}-\partial_\lambda\frac{\partial\mathcal{L}}{\partial
g_{\rho\sigma,\lambda}}+\partial^2_{\lambda\xi}\frac{\partial\mathcal{L}}{\partial
g_{\rho\sigma,\lambda\xi}}\biggr)\delta
g_{\rho\sigma}&=&\\\nonumber\doteq\int
d^4x\sqrt{-g}H^{\rho\sigma}\delta g_{\rho\sigma}&=&0.
\end{eqnarray}

The Euler-Lagrange
equations are then

\begin{eqnarray}
\frac{\partial \mathcal{L}}{\partial
g_{\rho\sigma}}-\partial_\lambda\frac{\partial\mathcal{L}}{\partial
g_{\rho\sigma,\lambda}}+\partial^
2_{\lambda\xi}\frac{\partial\mathcal{L}}{\partial
g_{\rho\sigma,\lambda\xi}}=0,
\end{eqnarray}
which coincide with the field Eqs. (\ref{fe1}) in  vacuum.
Actually, even in the case of  more general theories,  it is possible to
define an energy-momentum tensor that 
 turns out to be defined as follows\,:

\begin{eqnarray}
t^\lambda_\alpha=\frac{1}{\sqrt{-g}}\biggl[\biggl(\frac{\partial\mathcal{L}}{\partial
g_{\rho\sigma,\lambda}}-\partial_\xi\frac{\partial\mathcal
{L}}{\partial
g_{\rho\sigma,\lambda\xi}}\biggr)g_{\rho\sigma,\alpha}+\frac{\partial\mathcal
{L}}{\partial
g_{\rho\sigma,\lambda\xi}}g_{\rho\sigma,\xi\alpha}-\delta^\lambda_\alpha\mathcal{L}\biggr]\,.
\end{eqnarray}
This quantity, together with the energy-momentum tensor of matter
$T_{\mu\nu}$, satisfies a conservation law as required by the Bianchi identities. In fact, in presence of matter, one has
${\displaystyle H_{\mu\nu}\,=\,\displaystyle\frac{\chi}{2}T_{\mu\nu}}$,  and then

\begin{eqnarray}
(\sqrt{-g}t^\lambda_\alpha)_{,\lambda}=-\sqrt{-g}H^{\rho\sigma}g_{\rho\sigma,\alpha}=-\frac{\mathcal{X}}{2}\sqrt{-g}T^{\rho\sigma}
g_{\rho\sigma,\alpha}=-\mathcal{X}(\sqrt{-g}T^\lambda_\alpha)_{,\lambda}\,,
\end{eqnarray}
and, as a consequence,

\begin{eqnarray}
[\sqrt{-g}(t^\lambda_\alpha+\mathcal{X}T^\lambda_\alpha)]_{,\lambda}=0
\end{eqnarray}
that is  the conservation law given by the Bianchi identities.  We can now write  the
expression of the energy-momentum tensor $t^\lambda_\alpha$ in term of the gravity action $f(R)$ and its
 derivatives:

\begin{eqnarray}\label{tens-f(R)}
t^\lambda_\alpha=f'\biggl\{\biggl[\frac{\partial R}{\partial
g_{\rho\sigma,\lambda}}-\frac{1}{\sqrt{-g}}\partial_\xi\biggl(\sqrt{-g}
\frac{\partial R}{\partial
g_{\rho\sigma,\lambda\xi}}\biggr)\biggl]g_{\rho\sigma,\alpha}+\frac{\partial
R }{\partial
g_{\rho\sigma,\lambda\xi}}g_{\rho\sigma,\xi\alpha}\biggr\}-f''R_{,\xi}\frac{\partial
R }{\partial
g_{\rho\sigma,\lambda\xi}}g_{\rho\sigma,\alpha}-\delta^\lambda_\alpha\
f\,,
\end{eqnarray}
It is worth noticing that $t^\lambda_\alpha$ is a non-covariant
quantity in GR while its  generalization, in fourth order gravity,
turns out to satisfy the covariance prescription of standard tensors (see also \cite{HOG}). On the other hand, such an
expression reduces to  the Landau-Lifshitz
energy-momentum tensor of GR as soon as  $f(R)\,=\,R$, that is

\begin{eqnarray}
{t^\lambda_\alpha}_{|_{\text{GR}}}=\frac{1}{\sqrt{-g}}\biggl(\frac{\partial\mathcal{L}_{\text{GR}}}{\partial g_{\rho\sigma,\lambda}}g_{\rho\sigma
,\alpha}-\delta^\lambda_\alpha\mathcal{L}_{\text{GR}}\biggr)
\end{eqnarray}
where the GR Lagrangian has been considered in its effective form,
i.e. the symmetric part of the Ricci tensor, which effectively
 leads to the 
equations of motion, that is 
\begin{equation}
\mathcal{L}_{\text{GR}}=\sqrt{-g}g^{\mu\nu}(\Gamma^\rho_{\mu\sigma}\Gamma^\sigma_{\rho\nu}-\Gamma^\sigma_{\mu\nu}\Gamma^\rho_{\sigma\rho})\,.
\end{equation}
It is important to stress that the definition of the
energy-momentum tensor in GR and in $f(R)$-gravity are  different.
This discrepancy is due to the presence, in the second case,
of higher than second order differential terms  that  cannot be discarded  by means of a boundary
integration as it is done in GR. We have noticed  that the
effective Lagrangian of GR turns out to be the symmetric part of
the Ricci scalar since the second order terms, present in the
definition of $R$ , can be removed by means of integration by parts.

On the other hand, 
an analytic $f(R)$-Lagrangian can be recast,  at linear order,
as $f\sim f'_0R+\mathcal{F}(R)$, where the function
$\mathcal{F}$ satisfies the condition: $\lim_{R\rightarrow
0}\mathcal{F}\rightarrow R^2$. As a consequence, one can rewrite
the explicit expression of $t^\lambda_\alpha$ as\,:

\begin{eqnarray}\label{tensorf(R).1}
t^\lambda_\alpha=f'_0{t^\lambda_\alpha}_{|_{\text{GR}}}+\mathcal{F}'\biggl\{\biggl[\frac{\partial
R}{\partial g_{\rho
\sigma,\lambda}}-\frac{1}{\sqrt{-g}}\partial_\xi\biggl(\sqrt{-g}\frac{\partial
R}{\partial
g_{\rho\sigma,\lambda\xi}}\biggr)\biggl]g_{\rho\sigma,\alpha}+\frac{\partial
R }{\partial
g_{\rho\sigma,\lambda\xi}}g_{\rho\sigma,\xi\alpha}\biggr\}-\mathcal{F}''R_{,\xi}\frac{\partial
R }{\partial
g_{\rho\sigma,\lambda\xi}}g_{\rho\sigma,\alpha}-\delta^\lambda_\alpha
\mathcal{F}.
\end{eqnarray}
The general expression of the Ricci scalar, obtained by splitting
its linear ($R^*$) and quadratic ($\bar{R}$) parts  once a
perturbed metric (\ref{me}) is considered, is

\begin{eqnarray}\label{defRicciscalar}
R=g^{\mu\nu}(\Gamma^\rho_{\mu\nu,\rho}-\Gamma^\rho_{\mu\rho,\nu})+g^{\mu\nu}(\Gamma^{\rho}_{\sigma\rho}\Gamma^
{\sigma}_{\mu\nu}-\Gamma^{\sigma}_{\rho\mu}\Gamma^{\rho}
_{\nu\sigma})= R^*+\bar{R}\,,
\end{eqnarray}
(notice that $\mathcal{L}_{\text{GR}}=-\sqrt{-g}\bar{R}$).
In the case of GR ${t^\lambda_\alpha}_{|_{\text{GR}}}$,
the Landau-Lifshitz tensor presents a first non-vanishing term at order
 $h^2$. A similar result can be obtained in the case of $f(R)$-gravity.  In fact, taking into account Eq.(\ref{tensorf(R).1}), 
 one obtains that, at the lower order,
$t^\lambda_\alpha$ reads\,:

\begin{eqnarray}\label{tensorf(R).2}
t^\lambda_\alpha\sim{t^\lambda_\alpha}_{|h^2}&=&f'_0{t^\lambda_\alpha}_{|_{\text{GR}}}+f''_0R^*\biggl[\biggl(-
\partial_\xi\frac{\partial R^*}{\partial
g_{\rho\sigma,\lambda\xi}}\biggr)g_{\rho\sigma,\alpha}+\frac{\partial
R^*}{\partial
g_{\rho\sigma,\lambda\xi}}g_{\rho\sigma,\xi\alpha}\biggr]-f''_0R^*_{,\xi}\frac{\partial
R^*}{\partial
g_{\rho\sigma,\lambda\xi}}g_{\rho\sigma,\alpha}-\frac{1}{2}f''_0\delta^\lambda_\alpha
{R^*}^2=\nonumber\\
&=&f'_0{t^\lambda_\alpha}_{|_{\text{GR}}}+f''_0\biggl[R^*\biggl(\frac{\partial
R^*}{\partial
g_{\rho\sigma,\lambda\xi}}g_{\rho\sigma,\xi\alpha}-\frac{1}{2}R^*\delta^\lambda_\alpha\biggr)-\partial_\xi\biggl(R^*\frac{\partial
R^*}{\partial
g_{\rho\sigma,\lambda\xi}}\biggr)g_{\rho\sigma,\alpha}\biggr]\,.
\end{eqnarray}
Considering the perturbed metric (\ref{me}), we have $R^*\sim R^{(1)}$,
where $R^{(1)}$ is defined as in (\ref{approx1}). In terms of $h$ and $\eta$, we get

\begin{eqnarray}
\left\{\begin{array}{ll}\frac{\partial R^*}{\partial
g_{\rho\sigma,\lambda\xi}}\sim\frac{\partial R^{(1)}}{\partial
h_{\rho\sigma,\lambda\xi}}=\eta^{\rho\lambda}\eta^{\sigma\xi}-\eta^{\lambda\xi}\eta^{\rho\sigma}\\\\\frac{\partial
R^*}{\partial
g_{\rho\sigma,\lambda\xi}}g_{\rho\sigma,\xi\alpha}\sim
h^{\lambda\xi}_{\,\,\,\,\,\,,\xi\alpha}-h^{,\lambda}_{\,\,\,\,\,\alpha}
\end{array}\right.\,.
\end{eqnarray}
Clearly, the first significant term in Eq. (\ref{tensorf(R).2}) is of
second order in the perturbation expansion. We can now write  the
expression of the energy-momentum tensor explicitly in term of the
perturbation $h$; it is 

\begin{eqnarray}
t^\lambda_\alpha&\sim&f'_0{t^\lambda_\alpha}_{|_{\text{GR}}}+f''_0\{(h^{\rho\sigma}_{\,\,\,\,\,\,\,,\rho\sigma}-\Box
h)[h^{\lambda\xi}_{\,
\,\,\,\,\,\,,\xi\alpha}-h^{,\lambda}_{\,\,\,\,\,\,\,\alpha}-\frac{1}{2}\delta^\lambda_\alpha(h^{\rho\sigma}_{\,\,\,\,\,\,\,,
\rho\sigma}-\Box
h)]\nonumber\\&&-h^{\rho\sigma}_{\,\,\,\,\,\,\,,\rho\sigma\xi}h^{\lambda\xi}_{\,\,\,\,\,\,\,,\alpha}+h^{\rho\sigma\,\,\,\,\,\,\,\,\,\,\lambda}_
{\,\,\,\,\,\,\,,\rho\sigma}h_{,\alpha}+h^{\lambda\xi}_{\,\,\,\,\,\,\,,\alpha}\Box
h_{,\xi}-\Box h^{,\lambda}h_{,\alpha}\}\,.
\end{eqnarray}
Considering the tilded perturbation metric $\tilde{h}_{\mu\nu}$, the more compact form

\begin{eqnarray}
{t^\lambda_\alpha}_{|_f}=\frac{1}{2}\biggl[\frac{1}{2}\tilde{h}^{,\lambda}_{\,\,\,\,\alpha}\Box\tilde{h}-\frac{1}{2}\tilde{h}_{,\alpha}
\Box\tilde{h}^{,\lambda}-\tilde{h}^{\lambda}_{\,\,\,\,\,\sigma,\alpha}\Box\tilde{h}^{,\sigma}-\frac{1}{4}(\Box\tilde{h})^2\delta^
\lambda_\alpha\biggr]\,,
\end{eqnarray}
is achieved.
As matter of facts, the energy-momentum tensor of the gravitational
field, which expresses the energy transport  during
the propagation, has a natural generalization in the case of
$f(R)$-gravity. We have adopted here the
Landau-Lifshitz definition but other approaches can be taken into account  \cite{multamaki}. The general definition of
${t^\lambda_\alpha}$, obtained above, consists of a
sum of  a GR contribution plus a term coming from 
$f(R)$-gravity\,:

\begin{eqnarray}
t^\lambda_\alpha=f'_0{t^\lambda_\alpha}_{|_{\text{GR}}}+f''_0{t^\lambda_\alpha}_{|_f}\,.
\end{eqnarray}
However,  as soon as  $f(R)=R$,  we obtains
$t^\lambda_\alpha={t^\lambda_\alpha}_{|_{\text{GR}}}$. 
As a final remark, it is worth noticing  that massive modes of gravitational field come out from  ${{t^\lambda_\alpha}}_{|_f}$ since $\Box\tilde{h}$ can be considered an effective scalar field moving in a potential: $t^\lambda_\alpha$, in this case,  represents a transport tensor.

\section{Concluding remarks}
\label{5}
In this paper, we have formally studied the post-Minkowskian limit of $f(R)$-gravity developing all the calculations in the Jordan frame. The main result is that, beside standard massless modes of GR, further massive modes emerge and they are directly determined by the analytic parameters of  $f(R)$-gravity, that is the coefficients $f'_0$ and $f''_0$ of the Taylor expansion. This fact is extremely relevant since it does not depend on the considered $f(R)$-model  but it  is a general feature that  can be enounciated in the following way: {\it Massless gravitons are a peculiar characteristic of GR while extended or alternative theories have, in general, further massive or ghost states} \cite{greci}.  It is worth noticing that several indications in this sense are present in literature \cite{stelle78, starobinsky80, kerner} but  their relevance, from an experimental viewpoint,  has never been stressed enough.

On the other hand, a similar result comes out also in the Newtonian limit of the same theories: Yukawa-like corrections to the gravitational potential emerge in general and they are absent only in the case of GR. It is interesting to note that also  the characteristic lengths of such corrections are related to  $f'_0$ and $f''_0$ as shown in \cite{noi-newt}. Also in this case, the Newtonian potential, coming from the weak field limit of GR, is only a particular case. 

These results pose interesting problems related to the validity of GR at all scales. It seems that it works very well at local scales (Solar System) where effects of further gravitational degrees of freedom cannot be detected. As soon as one is investigating larger scales, as  those of  galaxies,  clusters of galaxies, etc.,  further corrections have to be introduced in order to explain both astrophysical large-scale dynamics \cite{noi-mnras,saffari} and cosmic evolution \cite{f(R)-noi,pogosian}.  Alternatively, huge amounts of dark matter and dark energy have to be invoked to explain the phenomenology, but, up today there are no final answer for these new constituents at fundamental level.
 Furthermore, the fact that, up to now, only massless gravitational waves have been investigated could be a shortcoming preventing the possibility to find out other forms of gravitational radiation. Tests in this sense could come, for example, from the stochastic background of gravitational waves  where massive modes could play a crucial role in the cosmic background spectrum \cite{bellucci,felix}.

\end{document}